\documentclass[pra,twocolumn,superscriptaddress,showpacs,preprintnumbers,amsmath,amssymb,floatfix]{revtex4}
\usepackage{graphicx}
\usepackage{dcolumn}% Align table columns on decimal point
\usepackage{bm}% bold math
\usepackage{latexsym,epsfig}
\usepackage{textcomp}
\usepackage[breaklinks=true, colorlinks, citecolor=blue, linkcolor=blue, urlcolor=blue]{hyperref}

\begin{document}

\title{Quasi-adiabatic dynamics of ultracold bosonic atoms in a one-dimensional optical superlattice}

\author{A. Dhar}
%\affiliation{Indian Institute of Astrophysics, II Block, Koramangala, Bangalore 560 034, India}
\affiliation{COMP Centre of Excellence and Department of Applied Physics, Aalto University, FI-00076 Aalto, Finland}

\author{D. Rossini} 
\affiliation{NEST, Scuola Normale Superiore and Istituto Nanoscienze-CNR, I-56126 Pisa, Italy}

\author{B. P. Das}
\affiliation{Indian Institute of Astrophysics, II Block, Koramangala, Bangalore 560 034, India}

\begin{abstract}
  We study the quasi-adiabatic dynamics for a one-dimensional system
  of ultracold bosonic atoms loaded in an optical superlattice. 
  Focusing on a slow linear variation in time of the superlattice potential,
  the system is driven from a conventional Mott insulator phase,
  to a superlattice-induced Mott insulator, crossing in-between a gapless critical
  superfluid region. Due to the presence of a gapless region, a number of defects 
  depending on the velocity of the quench appear. Our findings suggest a 
  power-law dependence similar to the Kibble-Zurek mechanism for intermediate 
  values of the quench rate.
  For the temporal ranges of the quench dynamics that we considered, 
  the scaling of defects depends non trivially on the width of the 
  superfluid region.
\end{abstract}


\maketitle

\section{Introduction}

Ultracold atomic gases in optical lattices provide a unique platform to probe 
a wide range of quantum phenomena with a high degree of controllability. 
The success of the Bose-Hubbard (BH) Hamiltonian~\cite{Fisher_1989, Jaksch_1998}
in elucidating the behavior of ultracold bosons in optical lattices has
stimulated a great deal of interest, from theoretical and experimental 
points of view. In particular, the seminal work on the Mott insulator (MI) to superfluid (SF)
quantum phase transition (QPT)~\cite{Greiner_2002} paved the way for a number of studies 
that led to the observation of various exotic 
quantum phases~\cite{OptLatt_Adv, ColdAtoms_RMP}. %, Esslinger_Science} 

Non-equilibrium quantum effects in such systems can be investigated
by varying in time parameters such as the optical lattice depth 
or the magnetic field close to a Feshbach resonance. 
Such experimental possibilities have spurred a renewed interest in the study 
of both sudden and quasi-adiabatic quenches~\cite{Dynamics_Rev1, Dynamics_Rev2}. 
The latter would provide important insights into non-equilibrium quantum phase transitions. 
In the presence of a ground-state energy gap always being finite during 
a very slow time evolution, the adiabatic theorem guarantees that the system will remain 
in the instantaneous ground state of the corresponding time-dependent Hamiltonian. 
However if a gapless region is crossed, 
the system will be unable to stay in its equilibrium ground state, 
regardless of how slowly it is quenched. 
The non-adiabatic evolution inevitably excites the system and a number of defects 
will appear in the evolved state. The mechanism of such defects formation had been first 
addressed by Kibble and Zurek (KZ) in the context of classical phase transitions 
in the early universe~\cite{Kibble_1976, Zurek_1985}, 
and more recently extended to the quantum regime for the case of adiabatic quenches 
across a single quantum critical point~\cite{Dorner_2005, Polkovnikov_2005}. 

The possibility to apply this kind of quenches has led 
to a number of theoretical studies addressing different types of 
many-body systems, including spin chains and ultracold quantum gases 
(see, e.g., Refs.~\cite{Clark_2004, Dziarmaga_2005, Schutzhold_2006, Cherng_2006, Cucchietti_2007, Cincio_2007, Caneva_2007, Caneva_2008, Pellegrini_2008, Sengupta_2008, Deng_2008, Divakaran_2008, Polkovnikov_2008, Degrandi_2008, Canovi_2009, Mukherjee_2010, Zimmer_2010, Trefzger_2011, Kollath_2011, DeMarco_2011, Kollath_2014, Dziarmaga_2014}).
Despite the large body of work in this field, several aspects involving the response 
of such systems to slow quenches have not been completely understood and deserve 
further investigation. It is believed that, in presence of non-isolated critical points
or of extended critical regions, the validity of the KZ mechanism is a priori not 
obvious, even if in some cases it is still possible to predict the defect density 
by identifying a dominant critical point, 
or by using scaling arguments~\cite{Pellegrini_2008, Deng_2008, Divakaran_2008}.

The dynamics of ultracold bosons in an optical lattice subjected to a quasi-adiabatic quench
has been theoretically analyzed for the MI-SF as well as the reverse transition~\cite{Clark_2004, Schutzhold_2006}. 
The SF-MI transition for a slow quench has been also studied taking the effects of the parabolic 
trapping potential into account~\cite{Kollath_2011}. 
These results highlighted the emergence of a scaling behavior for the characteristic
length scale as a function of the quench rate, which is well approximated by a power law. 
However it has been later shown that, for a phase transition of the Kosterlitz-Thouless type 
(as is the case for the MI-SF transition in the one-dimensional BH model),
the exponents depend on that rate, and are generally different from the KZ prediction, 
based on the critical exponents that are relevant for asymptotically long quench times~\cite{Dziarmaga_2014}.
Experimental evidence in support of the growth 
of the condensate excitations with a power-law dependence on the quench rate has been observed 
for the MI-SF transition for ultracold bosons in an optical lattice~\cite{DeMarco_2011}. 
Furthermore, for the same transition in a similar system, the observation of the emergence 
of coherence and a power-law dependence of the correlation length on the quench rate 
for intermediate quenches has recently been reported~\cite{Bloch2014}.

The feasibility of superposing different optical lattices with distinct 
frequencies~\cite{Folling_2007} also made the study of local relaxation dynamics 
possible in such superlattice setups~\cite{Cramer_2008, Trotzky_2011}.
An interesting property of these composite structures when they are loaded with 
bosonic atoms is that they can facilitate the generation of multiple lattice-modulated MI phases, 
which can isolate SF regions in the parameter space of the system~\cite{DharMFT, Dhar}. 
To the best of our knowledge, slow quenches for QPTs in optical superlattices starting 
and ending in insulating phases, and crossing a superfluid region in between, 
have not been addressed so far, and this is the focus of our present work.

Here we consider a one-dimensional (1D) system of ultracold bosonic atoms 
loaded in an optical superlattice, 
formed by two superimposed optical standing waves with different frequencies.
At zero temperature, this system exhibits different quantum phases: MI, SF and a 
superlattice-induced MI (SLMI) with periodically modulated onsite occupation~\cite{DharMFT, Dhar}.
The superlattice potential is constructed as to vary linearly in time, and
is chosen in such a way that it crosses a gapless region in between two insulating regions. 
We consider the formation of defects in the final state after the quench, 
and demonstrate a non-trivial scaling of the excess energy as a function of the quench rate. 
We tackle this problem by means of the time-dependent density matrix renormalization group (DMRG) 
method, in the formalism of matrix-product-state ansatz~\cite{Schollwock_2011}.

The paper is organized as follows. We start introducing our model and
discussing the static properties of its ground-state phase diagram that are relevant 
to the ongoing discussion (Sec.~\ref{sec:model}).
In Sec.~\ref{sec:quench} we define our dynamical protocol and discuss the formation of defects 
and the behavior of two-point correlation functions at the end of the protocol.
Finally, in Sec.~\ref{sec:summary} we draw our conclusions.

\section{Model and phase diagram} 
\label{sec:model}

\begin{figure}[t]
  \includegraphics[width=0.8\linewidth]{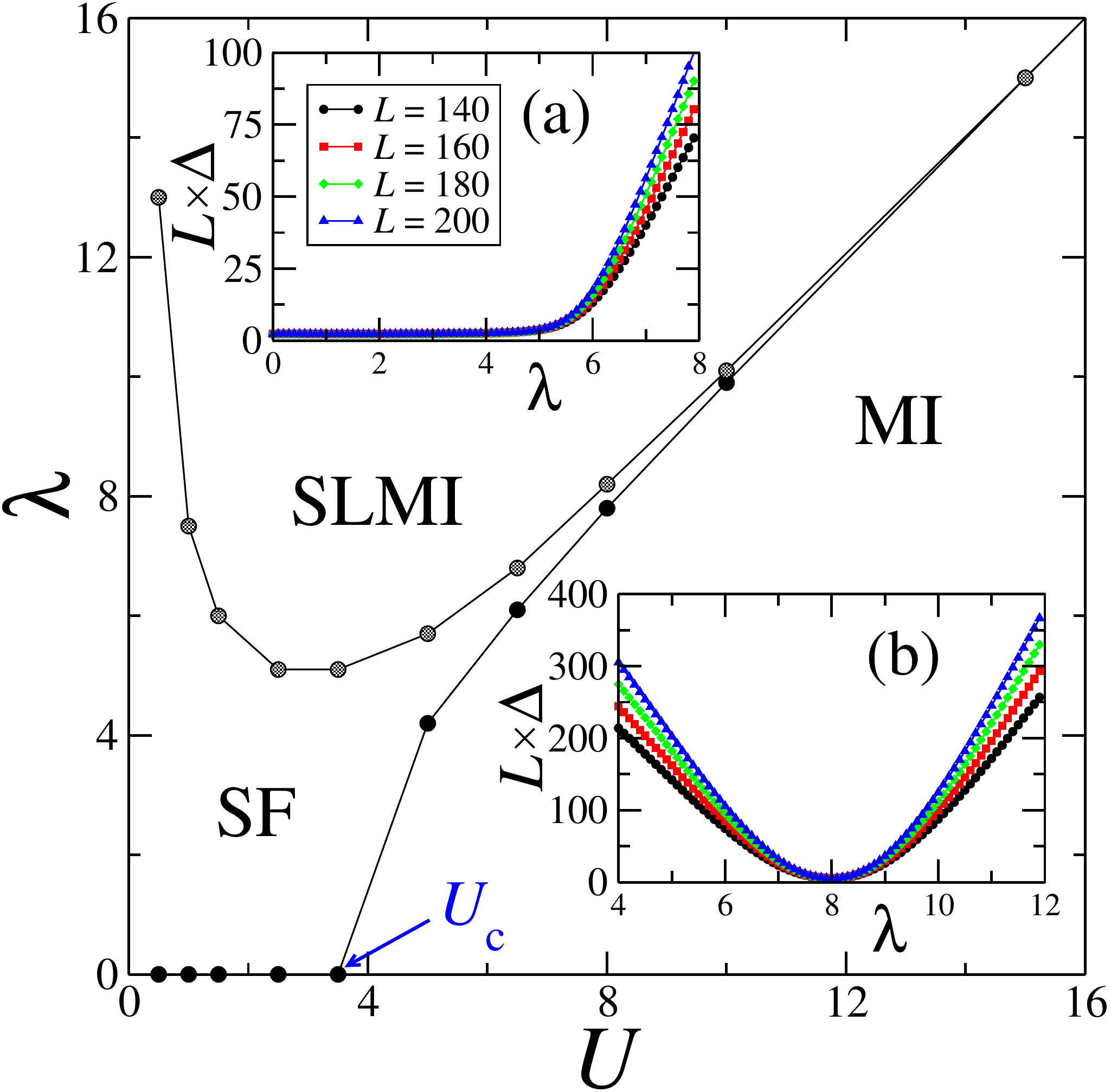} 
  \caption{(color online). Zero-temperature phase diagram in the $\lambda$-$U$ plane
    of the Hamiltonian $\hat{\cal H}_{\rm SLBH}$ for a period-two optical superlattice
    and with unitary filling. 
    Data have been obtained from DMRG calculations of the charge gap,
    that is the energy for adding or removing a particle from the state with $\bar{n}=1$.
    Insets: Ground-state energy gaps as function of $\lambda$ and for different system sizes 
    and two values of $U$, {\bf (a)} smaller than $U_c$ ($U=2.5$), and {\bf (b)} larger than $U_c$ ($U=8.0$).
    The gaps in the SF region close as the inverse system size $L^{-1}$.
    Here and in the subsequent figures the Hamiltonian parameters $\lambda$ and $U$ as well as
    the energy gaps $\Delta$ are expressed in units of $J$.}
  \label{fig:phase_diagram}
 \end{figure}

The model is described by the following Hamiltonian: 
\begin{equation}
  \hat{\cal H}_{\rm SLBH} = \sum_i -J (\hat{a}_{i}^{\dagger}\hat{a}_{i+1} + \mbox{H.c.})
  + \frac{U}{2} \hat{n}_{i}(\hat{n}_{i}-1) + \lambda_{i}\hat{n}_{i}\,,
  \label{eq:SLBHmodel}
\end{equation}
where $\hat{a}^\dagger_i, \hat{a}_i$ denote the creation and annihilation operators 
on site $i$ satisfying the usual bosonic commutation rules, with 
$\hat{n}_i = \hat{a}^\dagger_i \hat{a}_i$ being the corresponding number operator. 
The parameter $J$ denotes the hopping amplitude, $U$ is the on-site repulsive interaction
strength and $\lambda_{i}$ quantifies the superlattice potential depth.
For the period-two optical superlattice that we have considered, 
$\lambda_{i}$ has a finite value of $\lambda>0$ for odd sites, and it is zero for even sites. 
Hereafter we work in units of $\hbar = 1$ and set $J=1$ as the energy scale.
The critical $U$ value for the MI-SF transition is located at $U_c \approx 3.3$ 
for $\lambda=0$~\cite{Kuhner_2000}, as shown in Fig.~\ref{fig:phase_diagram}.

%%%%%%%%%%%%%%%%%%%%%%%%%%%%%%%%%%%%%%%%%%%%%%%%%%%%%%%%%%%%%%%%%%%%%%%%%%%%%%%%%%%%%%%%%
%%%%%%%%%%%%%%%%%%%%%%%%%%%%%%%%%%%%%%%%%%%%%%%%%%%%%%%%%%%%%%%%%%%%%%%%%%%%%%%%%%%%%%%%%

The ground-state phase diagram of model~\eqref{eq:SLBHmodel} has been studied 
by means of mean field theory~\cite{DharMFT}, quantum Monte Carlo techniques~\cite{Rousseau_2006} 
and the DMRG method~\cite{Dhar}. In 1D and at integer filling $\bar{n} = 1$,
this is given in the $\lambda$-$U$ plane as in Fig.~\ref{fig:phase_diagram}. 
Here we identify the various phases by analyzing the behavior of the ground-state 
energy gap $\Delta$ as a function of the system size in the following way. 
First we observe that, for a MI, this is finite and coincides with 
the charge gap $\Delta_+ - \Delta-$, that is the difference between the energy cost 
to add ($\Delta_+$) and to remove ($\Delta_-$) a particle from the system.
On a chain of finite length $L$, the numerical evaluation of the Mott gap 
has been thus obtained by performing three DMRG iterations, 
with projections on different number sectors $L, \, L \pm 1$. The corresponding 
ground states respectively give the desired energies $E_0$ , $E_\pm = E_0 + \Delta_\pm$.
In the SF region, this gap vanishes as the inverse of the system size $L$. 
The critical points have been extracted as those in which the product $\Delta \times L$
for the smallest ($L = 140$) and the largest ($L = 200$) considered lengths 
differed more than four percent (see the insets of Fig.~\ref{fig:phase_diagram}).

The nature of the insulating phase (MI or SLMI) depends on the relative strength 
of $U$ and $\lambda$. A SF region is present in between these two insulators, and its 
width decreases with increasing $U$. 
The intervening SF phase arises because of the competition between 
the superlattice potential $\lambda$ and the on-site two-body interaction $U$.
For $U < U_c$, there is only a transition from a gapless SF to a gapped SLMI
at a critical value of $\lambda$. But for $U > U_c$, there are three possible scenarios.
If $\lambda$ is much smaller than $U$, the system remains in the MI phase. 
When it becomes comparable to $U$, the system makes a transition to the SF phase, 
and for large values of $\lambda$ it enters the SLMI phase.

%%%%%%%%%%%%%%%%%%%%%%%%%%%%%%%%%%%%%%%%%%%%%%%%%%%%%%%%%%%%%%%%%%%%%%%%%%%%%%%%%%%%%%%%%
\begin{figure*}
  \includegraphics[width=0.48\linewidth]{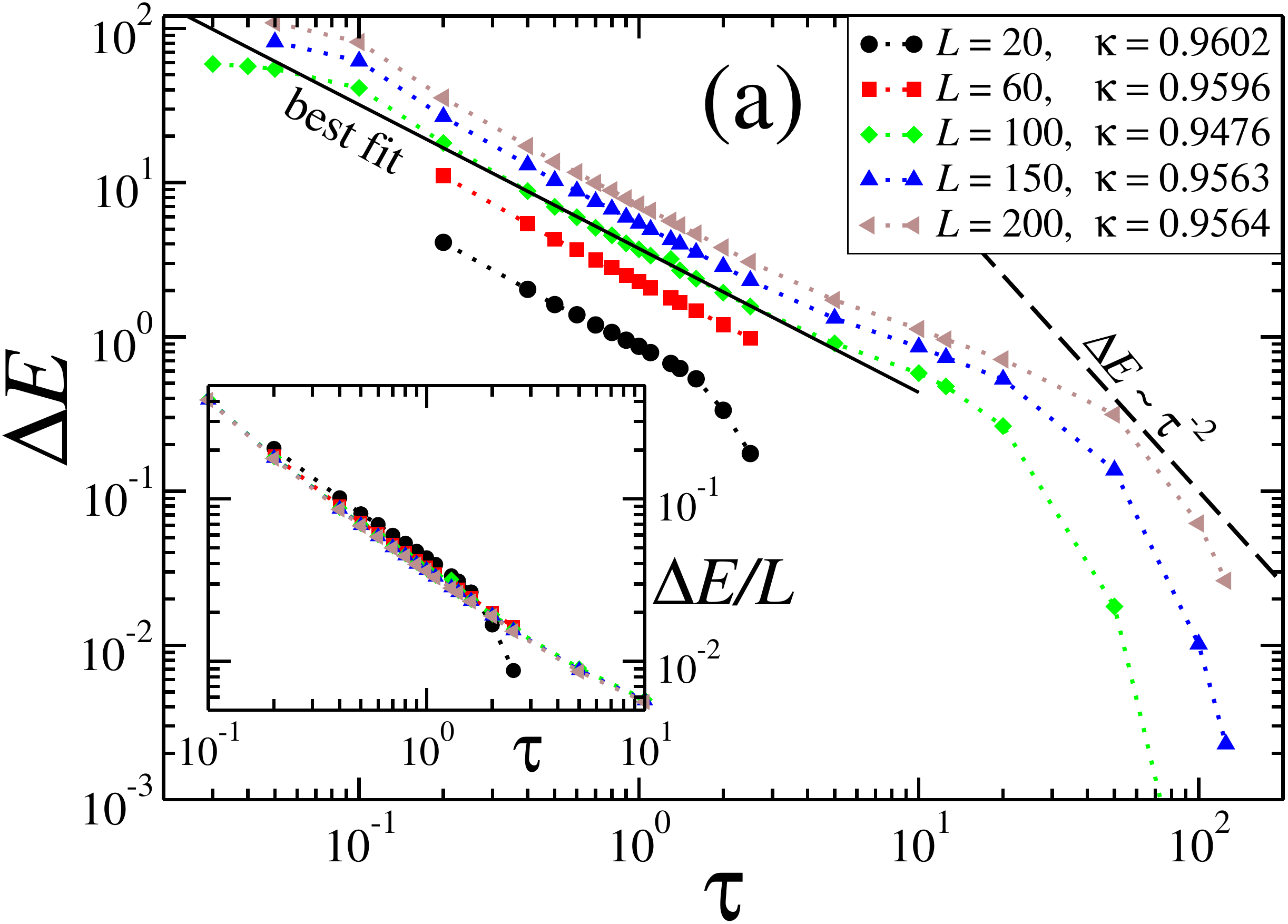} \hspace*{0.5cm}
  \includegraphics[width=0.48\linewidth]{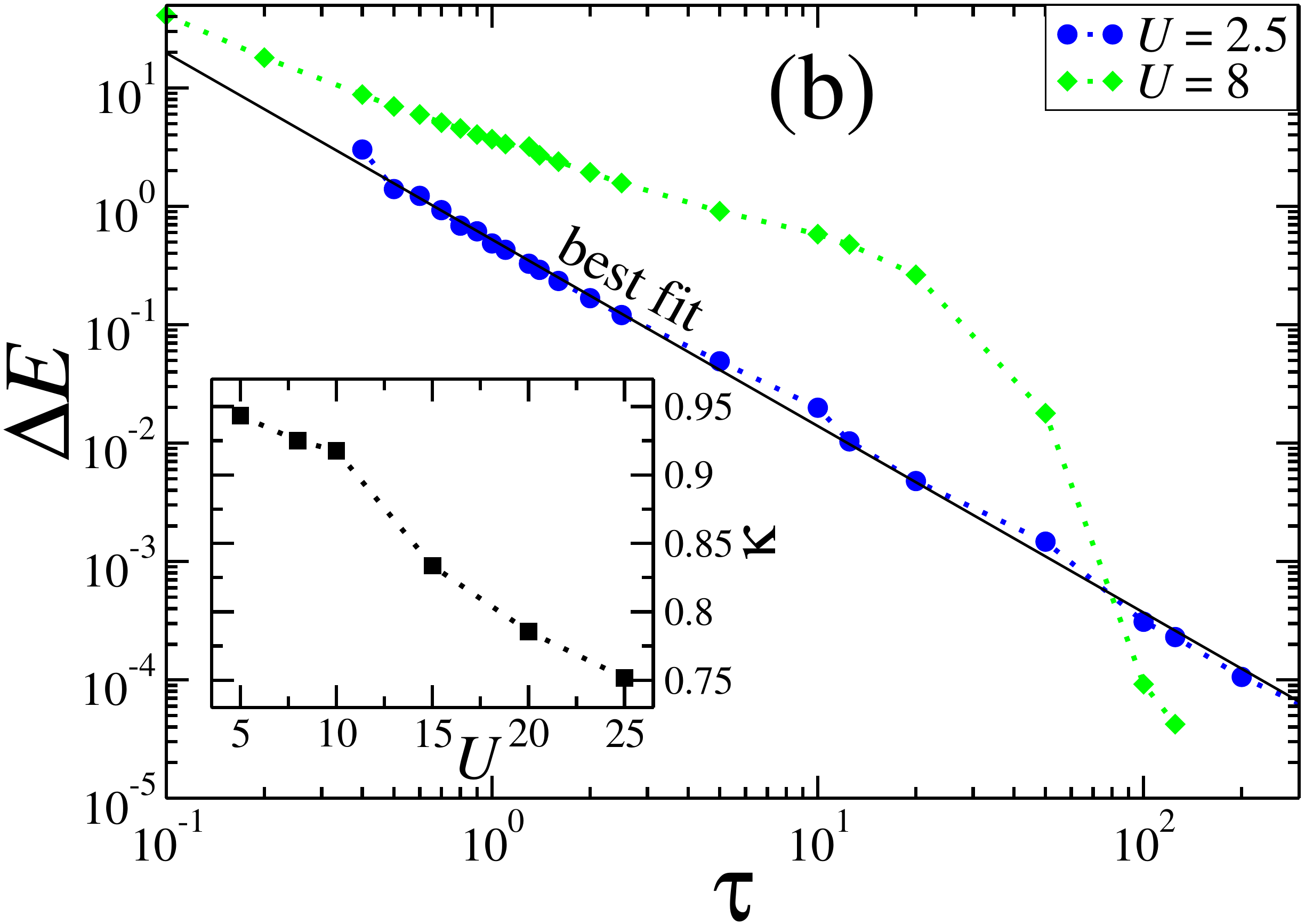} 
  \caption{(color online). Excess energy $\Delta E(\tau)$ at the end of the time evolution, 
    as a function of the quench time, $\tau$.
    Panel {\bf a)}: quench from MI to SLMI phase, with a SF phase in between.
    Here we fix $U=8$ and $\lambda: 5.5 \to 10.5$.
    The various data sets stand for different system lengths.
    Our fitting procedure in the intermediate region gives the values of $\kappa$ as shown in the legend.
    Inset: same data rescaled over the size $L$.
    Panel {\bf b)}: quench from SF to SLMI phase (blue circles: $U=2.5$, $\lambda: 0 \to 5$, $L=100$);
    the green diamonds set is the same as that reported in panel a).
    The continuous straight line is a power-law fit of the DMRG data, and gives a
    value of $\kappa \approx 1.5754$. 
    Inset: Power-law decay rate $\kappa$ for the excess energy 
    in the intermediate scaling region, as a function of the onsite interaction $U$,
    for quenches across the MI $\to$ SF $\to$ SLMI phase transitions ($U>U_c$).
    Here and in the subsequent figures times are expressed in units of $\hbar / J$,
    while excess energies $\Delta E$ are given in units of $J$.}
  \label{fig:DeltaE_Tau}
\end{figure*}
%%%%%%%%%%%%%%%%%%%%%%%%%%%%%%%%%%%%%%%%%%%%%%%%%%%%%%%%%%%%%%%%%%%%%%%%%%%%%%%%%%%%%%%%%

\section{Quasi-adiabatic quench dynamics}
\label{sec:quench}

In order to probe the slow quench dynamics of ultracold atoms in an optical superlattice, 
Eq.~\eqref{eq:SLBHmodel}, it is necessary to analyze the excitations 
that are generated when the gapless SF phase is crossed. 
In view of the specific features of the phase diagram, 
it is convenient to fix a value of $U$ and increase the parameter $\lambda$, so as to drive
the system across the MI-to-SF and then the SF-to-SLMI phase transitions.
As can be seen in Fig.~\ref{fig:phase_diagram}, the width of the critical region 
changes with $U$. This reflects into a non trivial dependence of the rate of defects 
generation with $U$, as discussed later.

We adopt a linear variation in time of the superlattice potential $\lambda$, 
which is given by
\begin{equation}
  \lambda(t) = \lambda_{0} + (\lambda_f -\lambda_0) \, t /\tau \,.
  \label{eq:lambdat}
\end{equation}
Here $\tau$ denotes the time of the quench, while $\lambda_{0}$ and $\lambda_f = \lambda(\tau)$ 
are respectively the initial and the final values of the superlattice potential. 
After fixing the value of $U$, we choose $\lambda_{0}$ and $\lambda_f$ such that 
the system starts from a MI and ends in a SLMI phase (except for the cases with $U < U_c$, 
where there is no MI phase). In between the initial and the final insulating phases,
there is a SF region, whose width depends on $U$. Due to the presence of a gapless 
region at the thermodynamic limit $L \to \infty$, a certain number of defects 
in the final state after the evolution will appear, 
no matter how slowly the quenching is performed~\cite{note1}. 
Below we shed light on these defects.

The system wavefunction $|\psi(t) \rangle$
evolves according to the time-dependent Schr\"{o}dinger equation. 
We computed $|\psi(\tau)\rangle$ at the final time $\tau$ after the quench~\eqref{eq:lambdat},
using a time-evolving block-decimation strategy~\cite{Schollwock_2011, Vidal_2004}.
We simulated systems up to $L=200$ sites with no more than $n_{\rm max} = 3$ bosons per site,
and used open boundary conditions.
The time interval $[0,\tau]$ has been discretized into many slices of time-length $\Delta t \ll 1$, 
where $\hat{\cal H}(t)$ is assumed to be constant.
The corresponding time evolution operator $e^{-i \hat{\cal H}(t) \Delta t}$
has been expanded by means of a sixth-order Suzuki-Trotter decomposition~\cite{Suzuki_Trotter}.
We have been able to consider a time-step $\Delta t = 0.05$, 
and reach a threshold for the discarded states $\varepsilon = 10^{-9}$,
by using a bond-link dimension $m \lesssim 200$ for all our simulations~\cite{note1b}.

\subsection{Excess energy}

To quantify the defects generation due to the non-adiabatic crossing of the SF region 
during the time evolution, we focus on the residual energy $\Delta E (t)$, 
defined as the excess energy above the ground state:
\begin{equation}
  \Delta E (t) = \langle \psi(t) | \hat{\cal H}(t) | \psi(t) \rangle 
  - \langle \psi_{\rm GS} (t) | \hat{\cal H}(t) | \psi_{\rm GS} (t) \rangle \,,
  \label{eq:ExcessEn}
\end{equation}
where $\langle \psi(t) | \hat{\cal H}(t) | \psi(t) \rangle$ denotes 
the energy of the system at time $t$, while 
$\langle \psi_{\rm GS} (t) | \hat{\cal H}(t) | \psi_{\rm GS} (t) \rangle$ 
is the instantaneous ground-state energy 
for Hamiltonian $\hat{\cal H}_{\rm SLBH}$ at time $t$.
This quantity serves as analog of the defect density originally considered 
by Kibble and Zurek (see, e.g., Refs.~\cite{Caneva_2007, Caneva_2008, Pellegrini_2008}).
Let us now discuss its behavior after a time modulation of the superlattice 
depth $\lambda(t)$, from $t=0$ to $t=\tau$, as dictated by Eq.~\eqref{eq:lambdat}.
In particular we focus on the scaling of $\Delta E (\tau)$ with $\tau$
for different values of interaction $U$.

The typical scenario is depicted in Fig.~\ref{fig:DeltaE_Tau}a,
where we are able to distinguish three distinct behaviors 
as a function of the quench time.
For $\tau \gg 1$, the dynamics is ruled by the adiabatic regime
typical of slow quenches: the time-evolved wavefunction remains very close 
to the instantaneous ground state of $\hat {\cal H} (t)$. 
The residual energy follows a power-law behavior 
\begin{equation}
  \Delta E \sim \tau^{-\kappa}
  \label{eq:Power-law}
\end{equation}
with $\kappa = 2$. 
This exponent can be obtained within the Landau-Zener formalism~\cite{Landau_Zener}, 
where the quantum evolution is studied by means of an effective two-level approximation 
with an avoided level crossing.
We point out that the adiabatic regime can occur only in the presence of an instantaneous 
ground-state energy gap which remains finite along the sweeping~\eqref{eq:lambdat} 
({\it i.e.}, for quenches much slower than the inverse square of the minimum crossed gap, 
$\tau \gg \Delta_{\rm min}^2$). Therefore this is a behavior related to a finite-size effect, 
which disappears at the thermodynamic limit where the gap in the superfluid region is rigorously zero.
In the opposite regime of fast quenches ($\tau \ll 1$), 
the dynamics is strongly non adiabatic and the initial state is essentially frozen
during the evolution. The excess energy thus saturates with $\tau$.
The intermediate regime in between is the most interesting one, since it is
crucially affected by the critical properties of the region crossed by the system.

In the intermediate regime, our data display a power-law scaling 
of the type in Eq.~\eqref{eq:Power-law}.
This fairly agrees with the general behavior predicted by KZ 
and verified in many cases, when the system is adiabatically driven
across isolated quantum critical points~\cite{Dynamics_Rev1}.
The KZ mechanism roughly identifies two types of evolution, 
either adiabatic or impulsive, according to the distance from the critical point. 
The time (distance from the critical point) at which the system switches 
to the impulsive regime depends on the quench velocity.
This simple argument indeed predicts a power-law scaling form for $\Delta E$
as a function of $\tau$, with a rate $\kappa$ expressed 
in terms of the critical exponents dictating the phase transition.
However, for the crossing of continuous phase transitions with extended 
critical regions, as it is in our case, the KZ scaling may still give insightful 
information, but cannot be regarded as ultimately predictive.
In the specific case of a Kosterlitz-Thouless transition 
analogous to that occurring in the BH model at $\lambda=0$, 
it has been shown that the exponentially slow gap closure induces a
power-law scaling which generally depends on the quench rate.
The exponent differs from that obtained with the usual KZ mechanism 
using the critical exponents of the transition~\cite{Dziarmaga_2014}.
In our specific situation, the system is quenched from a gapped to a gapless phase, 
and then to another gapped phase, thus crossing two QPTs and an extended 
critical (SF) region (Fig.~\ref{fig:phase_diagram}, for fixed $U > U_c$).
This is an even more complex scenario, where it is impossible to grasp
the quantitative power-law scaling predicted by KZ, and hence we expect 
the emergence of a more complex and inhomogeneous behavior, in terms of 
the size of the crossed critical region. 
%[[Besides that I am even not sure about the fact that the MI-SF transition
%at $\lambda \neq 0$ is of the Kosterlitz-Thouless type]]}

In the case of $U=8$ we clearly identify an intermediate scaling region 
with $\kappa \approx 0.95(1)$, as extracted from fits of the numerical data 
(see the solid line in Fig.~\ref{fig:DeltaE_Tau}a,
denoting the best fit of the data series at the size $L=100$).
Our results do not display a significant dependence of $\kappa$ on $L$.
Notice also that the excess energy per unit length is universal in the scaling region,
whose width increases with $L$, due to the gap closure 
in the SF phase (inset of Fig.~\ref{fig:DeltaE_Tau}a).
A qualitatively similar behavior is observed for a quench from SF to SLMI phase, 
at fixed $U < U_c$---Fig.~\ref{fig:DeltaE_Tau}b, blue circles, shows an example with $U=2.5$.
The value $\kappa \approx 1.58$ in the intermediate region obtained for that case 
is considerably larger than that for $U=8$.

%%%%%%%%%%%%%%%%%%%%%%%%%%%%%%%%%%%%%%%%%%%%%%%%%%%%%%%%%%%%%%%%%%%%%%%%%%%%%%%%%%%%%%%%%
\begin{figure}[!t]
  \includegraphics[width=0.9\linewidth]{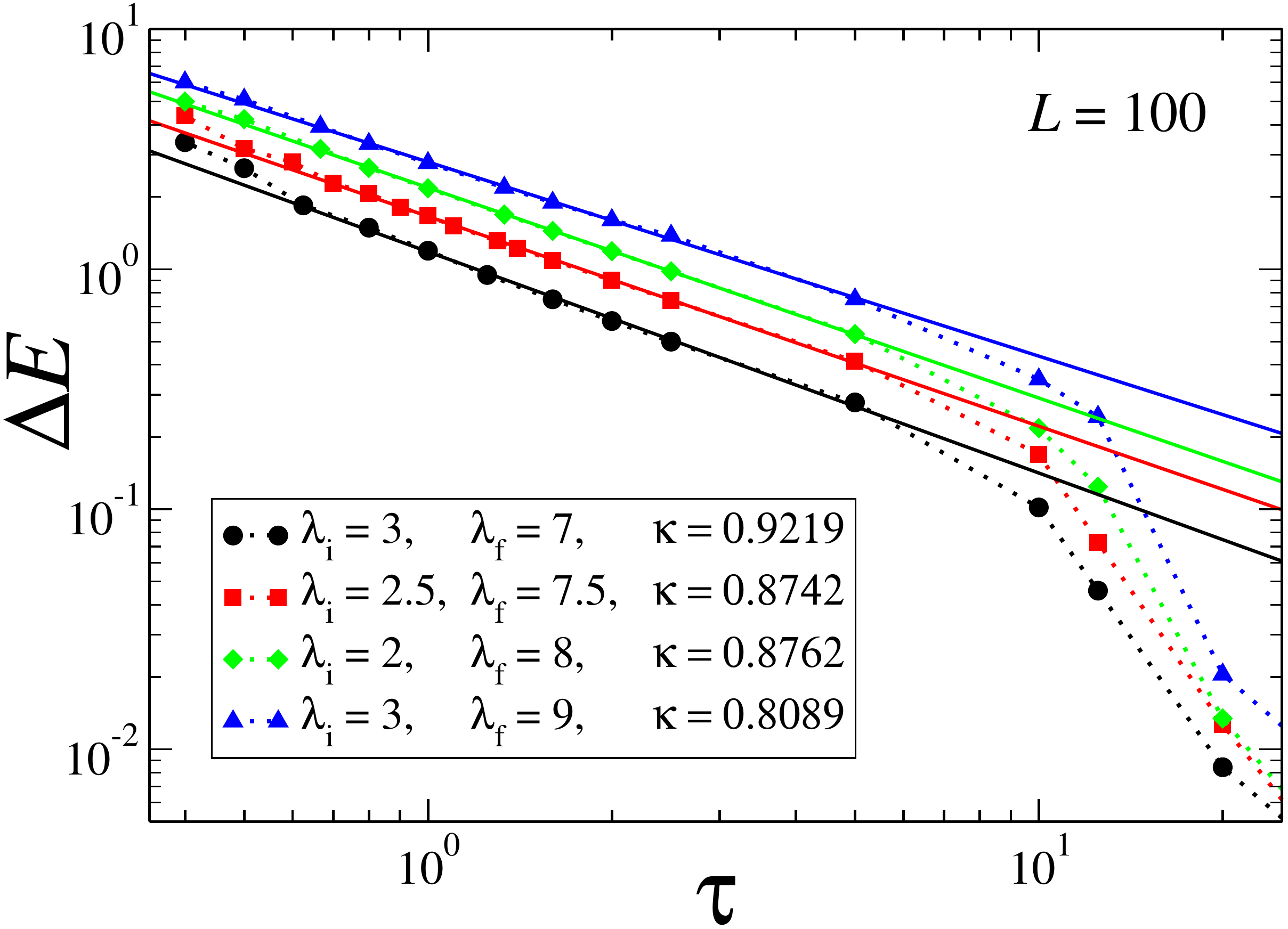} 
  \caption{(color online). Excess energy as a function of the quenching time for fixed $U=5$ 
    and a system size $L = 100$ for different initial and final points.}
  \label{fig:U5_LambdaVar}
\end{figure}
%%%%%%%%%%%%%%%%%%%%%%%%%%%%%%%%%%%%%%%%%%%%%%%%%%%%%%%%%%%%%%%%%%%%%%%%%%%%%%%%%%%%%%%%%

These observations support the evidence that any appropriate scaling analysis should depend 
non trivially on $U$, while a simple KZ argument cannot predict this complex behavior.
We point out that it is also not guaranteed that $\kappa$ does not change 
with the quench rate, as theoretically predicted for the MI-SF transition 
at $\lambda=0$~\cite{Dziarmaga_2014}.
In the range of $\tau$ we were able to address, we did not observe such dependence. 
However, as reported in the inset of Fig.~\ref{fig:DeltaE_Tau}b, 
for quenches across the phases MI $\to$ SF $\to$ SLMI with $U > U_c$, we found a rather complex dependence 
of $\kappa$ on $U$, and hence on the width of the intermediate SF region. 
In particular the defects production rate $\kappa$ decreases monotonically 
as a function of the time during which the system is crossing the gapless region.

We also checked the dependence of $\kappa$ on the starting and ending points
inside the insulator (we varied $\lambda_i$ and $\lambda_f$, for fixed $U$).
Results in Fig.~\ref{fig:U5_LambdaVar} indicate a tendency toward a slight decrease
of $\kappa$, if the gapped region crossed by the quench increases.

\subsection{Correlation functions}

\begin{figure}[!t]
  \includegraphics[width=0.9\linewidth]{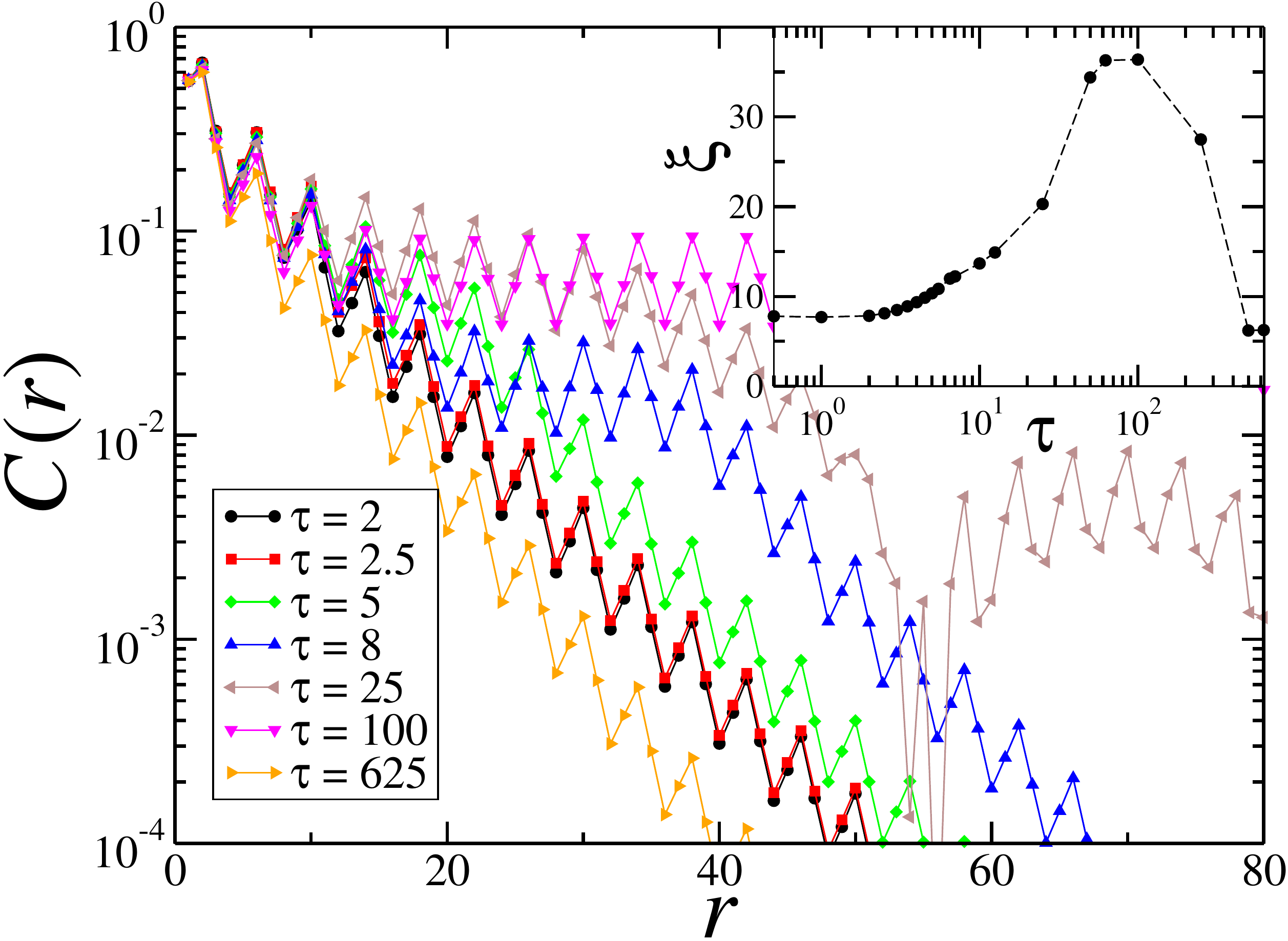} 
  \caption{(color online). Two-point correlation function 
    $C(r) = \langle \psi(\tau) \vert \hat a^\dagger_i \hat a_j \vert \psi(\tau) \rangle$
    as a function of the distance $r = \vert i-j \vert$ and for different quench times, $\tau$.
    The inset shows the correlation length $\xi$ as extracted from numerical data
    as a function of $\tau$. Data are for $U=8$ and $\lambda : 5.5 \to 10.5$. 
    Here we set $L = 100$. Distances are in units of the lattice spacing.}
  \label{fig:Corr_Length_U8}
\end{figure}

Finally we examined the behavior of the two-point correlation function 
$C(r) = \langle \psi(\tau) \vert \hat a^\dagger_i \hat a_j \vert \psi(\tau) \rangle$
at the end of the quasi-adiabatic dynamics, and observed an exponential decay 
with the distance $r = \vert i-j \vert$, as shown in Fig.~\ref{fig:Corr_Length_U8}.
The two points $i$ and $j$ have been chosen in a symmetric way with respect
to the center of the chain in order to minimize boundary effects, 
such that $i = (L - r + 1)/2$, $j = (L + r + 1)/2$ for odd $r$,
and $i = (L - r)/2$, $j = (L + r)/2$ for even $r$ (for instance, for $L = 100$ sites, 
$r = 1$ corresponds to $i = 50$, $j = 51$; $r = 2$ corresponds to $i = 49$, $j = 51$; 
$r = 3$ corresponds to $i = 49$, $j = 52$, and so on).

In the inset we plot the correlation function
\begin{equation}
  \xi = \sqrt{ \frac{\sum_r r^2 \langle \hat a_i^{\dagger} \hat a_j \rangle}{\sum_r \langle \hat a_i^{\dagger} \hat a_j \rangle}} , \qquad r = \vert i-j \vert \,.
  \label{eq:Corr}
\end{equation}
This clearly exhibits a non-monotonic behavior as a function of the quench rate $\tau$~\cite{note2}.
In particular we notice that $\xi(\tau)$ is increasing initially in the intermediate scaling region. 
This can be attributed to the persistence of quasi-long range order which the system gained 
while quenching through the SF phase. But after some critical value of $\tau$, we observe that 
$\xi$ starts to decrease. Such a behavior is ascribed to the onset of adiabatic regime. 
Since the system ends up in an insulating phase (SLMI), where the correlation function
decays exponentially, we expect in the adiabatic regime the correlation length to be small. 
In particular, the value of $\tau$ at which the transition from the intermediate regime to 
the adiabatic regime takes place, obtained from the analysis of the residual energy (Fig.~\ref{fig:DeltaE_Tau})
coincides with that seen in the correlation length (Fig.~\ref{fig:Corr_Length_U8}). 

%%%%%%%%%%%%%%%%%%%%%%%%%%%%%%%%%%%%%%%%%%%%%%%%%%%%%%%%%%%%%%%%%%%%%%%%%%%%%%%%%%%%%%%%%
%%%%%%%%%%%%%%%%%%%%%%%%%%%%%%%%%%%%%%%%%%%%%%%%%%%%%%%%%%%%%%%%%%%%%%%%%%%%%%%%%%%%%%%%%

\section{Summary}
\label{sec:summary}

We have theoretically analyzed the slow quench dynamics of ultracold bosonic
atoms in a one-dimensional optical superlattice. By considering a linear time dependence of
the superlattice potential, we showed that, when crossing
a gapless superfluid region, the system has the tendency to generate defects.
This fact is due to the adiabaticity loss during the time evolution, which occurs even when
the system is quenched very slowly. Our results show a complex dependence of the rate 
of defect generation on the quench velocity, which cannot be understood 
in terms of the Kibble-Zurek physics underlying the crossing of a single critical point. 

From an experimental point of view, the behavior of the excess energy
could be verified by means of time-of-flight measurements of the correlation length.
This, in turn, may reveal itself as a simple indicator of the presence or absence of the
power-law scaling regime for the defects production as a function of the quench velocity.
Trapping ultracold bosonic atoms in optical standing waves is probably 
the most favourable setup to probe this kind of physics. Recent experiments have 
already verified the power-law behavior~\cite{DeMarco_2011, Bloch2014} 
in the framework of the Bose-Hubbard model.
Moreover, the capability that has been demonstrated by a variety of 
out-of-equilibrium experiments with great accuracy and for large coherence times,
ranging from sudden quenches to adiabatic variation of the system's parameters,
puts our results arising from the quenching of
the superlattice potential in a favourable position for verification in the laboratory.

\acknowledgments

We thank R. Fazio for useful discussions. AD and BPD thank D. Sen and A. Dutta for stimulating 
discussions. AD acknowledges the Academy of Finland through its Center of Excellence Programme (2012-2017)
and under Project No.~13251748. DR acknowledges support from the Italian MIUR through FIRB Project RBFR12NLNA. 
The computations were performed on the Intel Cluster at the Indian Institute of Astrophysics.


\begin{thebibliography}{99}
\bibitem{note1}
  Note that, by first taking the thermodynamic limit $L \to \infty$, we are guaranteed that
  the system in the SF phase will be gapless, therefore the evolution will never be 
  adiabatic and defects will necessarily be formed at any finite value of $\tau$.
  On the contrary, by first taking the asymptotically slow quench limit $\tau \to \infty$,
  the evolution will stay adiabatic at any size $L$, and defects will not form.
\bibitem{note1b}
  We have checked that, on the scale of all the figures presented here, 
  our results do not present significant deviations, if we take 
  a time step $\Delta t = 0.025$ corresponding to half of the value used 
  throughout the simulations shown in this paper.
  This produces a discrepancy in the decay rates $\kappa$ 
  which turns out to be less than $0.5 \%$.
\end{thebibliography}
\end{document}